# A method for delineation of bone surfaces in photoacoustic computed tomography of the finger

**S. K. Biswas**[a,b*], **P. van Es**[a*], **W. Steenbergen**[a] **and S. Manohar**[a,†]

[a]Biomedical Photonic Imaging, University of Twente, P. O. Box 217,

7500 AE Enschede, The Netherlands.

[b]Department of Instrumentation and Applied Physics,

Indian Institute of Science, Bangalore 560012, India.

*Equal contribution

[†]**Address all correspondence to:** S.Manohar, University of Twente, TNW, Biomedical Photonic Imaging, Drienerlolaan 5, Enschede, The Netherlands, 7522DB; Tel: +31-(0)53-4893164; Fax: +31-(0)53-4891105; E-mail: s.manohar@utwente.nl

Key words: Photoacoustic tomography, Photoacoustically-induced ultrasound, Bone delineation, finger.

**Abstract**

Photoacoustic imaging of interphalangeal peripheral joints is of interest in the context of using the synovial membrane as a surrogate marker of rheumatoid arthritis. Previous work has shown that ultrasound produced by absorption of light at the epidermis reflects on the bone surfaces within the finger. When the reflected signals are backprojected in the region of interest, artifacts are produced, confounding interpretation of the images. In this work, we present an approach where the photoacoustic signals known to originate from the epidermis, are treated as virtual ultrasound transmitters, and a separate reconstruction is performed as in ultrasound reflection imaging. This allows us to identify the bone surfaces. Further, the identification of the joint space is important as this provides a landmark to localize a region-of-interest in seeking the inflamed synovial membrane. The ability to delineate bone surfaces allows us not only to identify the artifacts, but also to identify the interphalangeal joint space without recourse to new US hardware or a new measurement. We test the approach on phantoms and on a healthy human finger.

# 1 Introduction

Rheumatoid arthritis (RA) is a chronic inflammatory disease which affects synovial joints with a prevalence of up to 1% of the world's population.[1] The disease can be progressive and may be severely debilitating due to swelling, stiffness and pain in the joints of the limbs and neck. There is no cure for RA, however the disease can be successfully managed using anti-inflammatory agents, and immunosuppressive drugs.[2] However, therapies are expensive and there are several drawbacks in the methods used for diagnosis of RA and monitoring of therapies.

Currently, diagnosis is made based on a review of clinical symptoms and physical examination of the joints according to the DAS28 scale or one of its variants. Blood tests based on identifying abnormal antibodies such as rheumatoid factors are conducted.[3,4] Imaging of the joints is performed as a part of the diagnostic workup and only performed if there is substantial uncertainty around the diagnosis. However, conventional x-ray radiography is not able to visualize key features in the joint such as synovium, edema, cartilage etc. and is mainly applied for late stage disease monitoring. Conventional ultrasound (US) imaging allows visualization of pannus and edema in progressed RA.[5] For early RA, conventional US imaging is less suitable due to low soft tissue contrast between inflamed tissue and synovial tissues.[6]

RA is characterized by synovitis, or inflammation of the synovial membrane that lines the interior of the joint capsule. There are indications from several studies that synovitis is a sensitive marker of disease activity and severity in RA.[7,8] The inflamed condition is associated with abnormal increased vascularization in the synovial membrane or synovium due to vasodilation and/or angiogenesis. Imaging of synovial vascularity and thus synovitis is performed using Doppler ultrasound imaging and Magnetic Resonance Imaging (MRI).[9,10] However, these methods used have drawbacks: Doppler US lacks sensitivity to small blood vessels,[11] and MRI requires contrast agents and is expensive.

Evidently, there is a need for sensitive imaging modalities that perform accurate diagnosis at early stages and make possible monitoring of response to the expensive therapies. Photoacoustic (PA) imaging can visualize vasculature with high contrast and resolution, and can potentially image synovial vascularity and facilitate diagnosis and monitoring of RA.[12-14] The technique utilizes short nanosecond light pulses that lead to thermoelastic expansion and the emission of US from light absorbing structures, which are predominantly blood vessels in soft tissue These US waves that propagate to the boundary of tissue can be registered in time using US detectors and the signals are used to reconstruct a distribution map of the blood vessels.

Recent work from our group on imaging vasculature in the healthy finger[14] showed the feasibility of making detailed transversal cross-section images at resolutions as good as 100 μm. Along with the rich complexity of blood vessels in the finger, however, reflection artifacts from bone surfaces could also be seen (dashed arrow in Figure 1(a)). These were caused by the backscattering of strong PA signals at the surface of bone. These PA signals were initially generated at the epidermis and are indicated by the solid arrow in Figure 1(a).

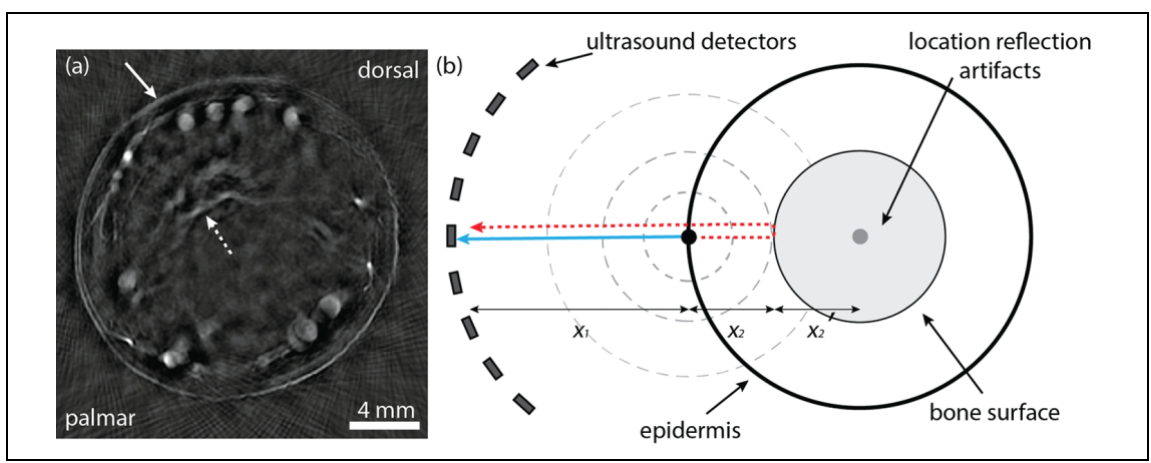

**Figure 1.** (a) Transverse PA slice image of index finger showing blood vessels (small circles and thread-like structures) and the epidermis (solid arrow). Acoustic artifacts (dashed arrow) can be seen, caused by acoustic backscattering from the bone of PA signals from the epidermis and blood vessel (reproduced from van Es et al. 2014 with permission from SPIE®). (b) Schematic representation of artifact generation in the case of PA finger imaging. PA signals generated by the epidermis travel to the detector directly (solid arrow) and after reflection at bone surface (dashed arrow). The reflected PA signal travels distance $x_1 + 2x_2$ from the epidermis to bone surface, and from bone surface to the detector. After PA reconstruction, the artifacts from the PA source are localized at distance $x_1 + 2x_2$ (gray dot). PA = photoacoustic.

The generation of these artifacts in PA images is depicted schematically in Figure 1(b). US from a PA source, such as the epidermis, is measured at the detector having travelled distance $x_1$ (solid arrow). The US will also have undergone reflection at the bone surface, having travelled distance $x_1 + 2x_2$ which is the sum of the distance from PA source to the bone and the distance from bone to the detector (dashed arrow). During reconstruction the reflected signals from the PA source will cause an artifact at a distance of $x_1 + 2x_2$ from the detector, shown as a grey dot in the figure.

In this case (Figure 1(a) and Ref. 14) the reflection artifacts were primarily visible within the bone volume and did not pose a problem. However these artifacts are physically unrealistic and could in other cases cause uncertainty in interpretation of PA images especially when vasculature close to the bone surface is present. Thus it is important to be able to discriminate between the PA signals and those from bone reflections. Further an interesting observation was made in our previous work[14], namely that the amount and/or intensity of these reflection artifacts was reduced at the estimated location of the joint space.[14] We believe that the location of the joint space in PA images is an important landmark, and together with some delineation of the bone will indicate the region-of-interest to seek the inflamed synovium.

In this work, we make the first step towards identifying bone-reflection artifacts, and present a method to recover the actual bone surface that causes these artifacts. The method considers the epidermis as a collection of ultrasound transmitters and the PA tomography probe as the detector array. It applies a pulse-echo algorithm to delineate surfaces of the finger bone and makes possible the localization of the joint space. By this method, no additional US transmitter hardware is required since the reflection maps are developed using photoacoustically-induced ultrasound (PAUS) signals. Due to the perfect match in time registration and US receiver locations, we are able to develop co-registered conventional PA and PAUS reflection images. We show feasibility of the method on finger phantoms and conclude by applying the concept to an *in vivo* measurement on a healthy finger.

## 2 Materials and methods

Imaging was performed using the PA computed tomography setup presented in Ref. 14. The object under investigation is held immobile in a water tank holding a 32-element curvilinear US array (Imasonic, Besançon, France) and 6 fiber bundles pointing towards the center of rotation (Figure 2(a)). The fiber bundles are evenly distributed in a half-circle around the object, have a 4 mm diameter and a NA of 0.22 in air. Six contiguous round spots of 11 mm with roughly top-hat intensity distribution at the finger or object surface are obtained. The 6 ns laser pulses are generated by an Nd:YAG laser (Quanta-Ray Pro 250, @10 Hz, Spectra Physics, Mountain View, CA) pumping an OPO (VersaScan-L532, GWU, Germany). The resulting PA signals are detected by the US detector array with a radius of curvature of 40 mm. The detector has a center frequency of 6.25 MHz and a -6 dB bandwidth of over 80%.[15] For data acquisition a 32-channel pulser/receiver (Lecoeur-Electronique, Chuelles, France) sampling at 80 MSs$^{-1}$ was used. The tank and its contents rotate in equiangular steps around 360° of the object providing a PA image slice, and then progressively down along the object to collect a stack of such slice images. The slice acquired has a -3 dB thickness of roughly 1 mm, due to focusing of the elements in the elevation plane at a distance of 48 mm. An in-plane resolution of 100 mm can be obtained from 12 views around the object using acoustic filtered backprojection.

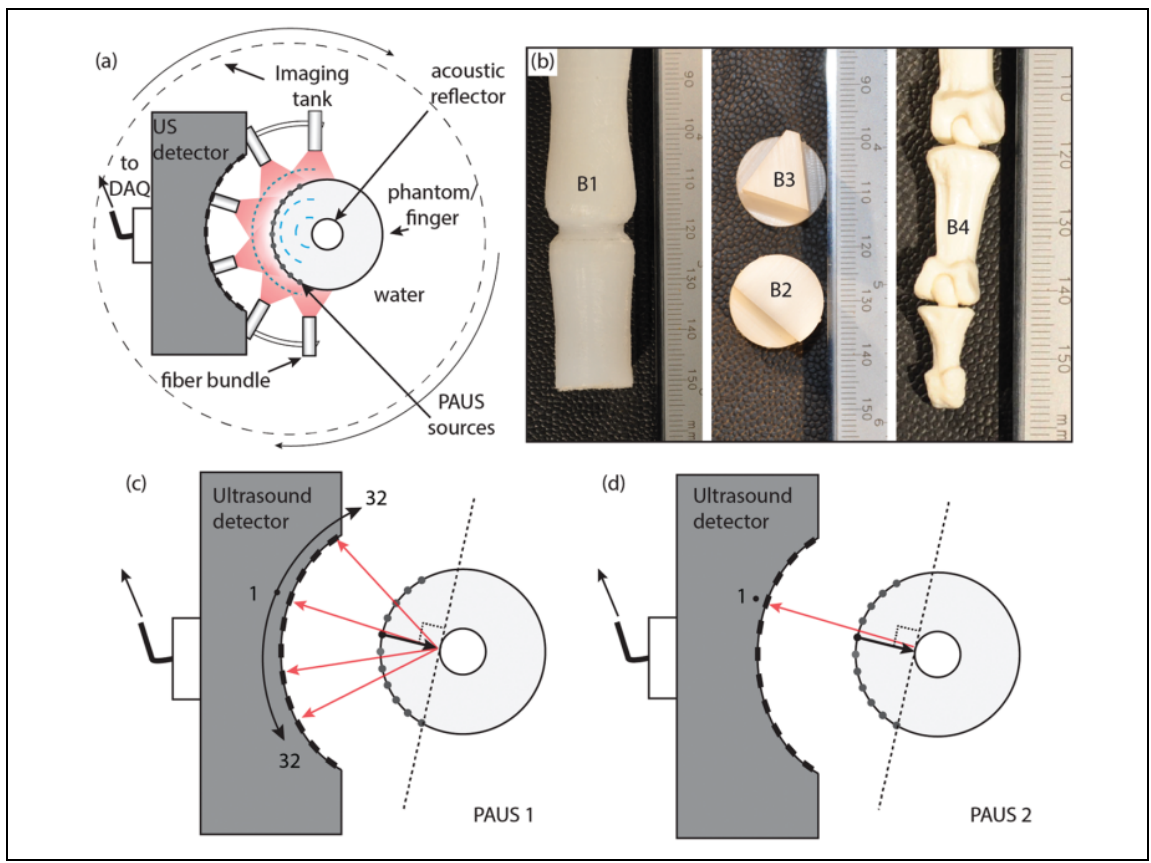

**Figure 2.** a) Schematic of the photoacoustic computed tomography setup. The detector and fiber bundles rotate around the finger or the agar gel finger phantom. (b) Photos of the bone-simulating acoustic reflectors of various materials and of various cross-sections. These were embedded in four different finger phantoms to test the feasibility of the approach in delineating the reflecting surfaces and identifying any clefts or joint spaces. (c) Schematic explanation of PAUS algorithm (PAUS-1). Each virtual pulser is active at a time; the scattered US from reflection at the bone surface is measured by 32 elements on either side of the acoustic axis of the pulser position. (d) Schematic explanation of PAUS algorithm (PAUS-2). Reconstruction with one pulser/receiver pair that are in one line toward the mechanical center of rotation of the setup. PAUS = photoacoustically induced ultrasound; US = ultrasound. DAQ = data acquisition card.

**Experimental phantom models**

To study the capability of the approach in delineating bone surfaces from the PAUS reflections, several phantom models were developed. From left to right in Figure 2(b), a rod with circular cross-section (B1), a rod with partly circular cross-section (B2), and a rod with cross-section in the shape of a trapezium (B3), were fabricated from Delrin®.

Further in order to study the ability of the method in localizing joint spaces, two models were chosen where the shape changes along the length of the rod. The model B1 (Figure 2(b) middle) was machined by lathe turning, so that its diameter varied along its length. The rod surface was slightly roughened with a file to improve US reflections in all directions. The second joint space containing model (B4) was made from the index finger of a skeleton hand model (Labor ActivA B.V, EZ6003, Steenbergen, The Netherlands). The overall shape of this bone model (Figure 2(b) right) is similar to real bone, apart from small rods in between the bones that serve as hinges at the joint space.

All rods were embedded in thin cylinders of 3% w/v agar gel formed using an aqueous solution of 1% Intralipid® (20%, Fresenius-Kabi Netherlands BV, Zeist, The Netherlands) to make finger phantoms, designated with letter P (P1-P4 respectively containing objects B1-B4) . The diameters of the phantoms are 26 mm (P1) or 20 mm (P2 and P3). For P4, the cross-section of the embedding agar was made roughly elliptical, by dipping the bone model several times into mobile agar gel at 40°C at various angles to obtain the approximate shape of a finger.

To simulate a thin melanin layer such as encountered in the epidermal skin layer, all the finger phantoms were dipped shortly into mobile 1% w/v agar gel at 40 °C containing 0.12% India Ink in water. Immediately after the phantoms were dipped in ice water (0 °C) to result in firm highly absorbing gel layers of 200-500 µm thickness.

## Phantom measurements

Phantoms P1, P2, and P3 were imaged using 12 view angles with 30 degrees spacing between each position. Phantom P4 was imaged using 24 view angles with 15 degrees spacing between each position. The imaging time per slice was approximately 60 seconds. A simple slice was acquired for the phantoms P2 and P3. Phantoms P1 and P4 were measured with respectively 60 and 100 slices along the whole length with an inter-slice spacing of respectively 1 mm and 0.5 mm. In these cases, the slice images were stacked to one dataset and smoothed and interpolated by a factor of 4 using the bicubic interpolation script in Matlab®. This compensated for differences between the resolution in the imaging plane and the resolution in the elevational plane.

### *In vivo* measurement

The index finger of a healthy volunteer with a dark skin type (Fitzpatrick scale skin Type V) was selected because of the high melanin content in the epidermis. This resulted in strong PA signals from melanin absorption at a wavelength of 700 nm while allowing sufficient light at deeper levels for visualization of blood vessels. The higher wavelength specific absorption in the epidermis at 700 nm made it unnecessary to differentiate between epidermal and blood vessel signals. At longer wavelengths highly absorbing blood vessels generate strong PA signals and can cause artifacts in the PAUS images. The imaging time per slice was approximately 60 seconds. Each of the 25 measured slices required a full 360 degrees rotation in 12 steps of 30 degrees. The fluence at the skin was approximately 6.5 mJ/cm$^2$ which is well below the maximum permissible exposure stated in IEC 60825- 1.

### Image reconstruction

The reflection artifacts in the finger are predominantly from those PA signals generated at the skin surface, by absorption at the epidermis, and reflection at the bone surface. We treat this situation as conventional (reflection mode) US imaging, where we assign to the epidermis an array of virtual ultrasound transmitters along the finger circumference that probe the finger tissue with US waves. These PAUS waves are detected by the ultrasound detectors after reflection on the bone surface.

For PA reconstruction, a standard algorithm based on acoustic filtered backprojection is used that projects the measured intensities from the ultrasound detector directly to the discretized absorption map in half-circles. Before backprojection, the PA data is filtered by a ramp filter and de-convolved with the system response function measured from a PA point source[15].

For reconstruction of the US reflections, the PAUS algorithm begins by identifying the locations of the epidermis in the finger or ink layers in the phantom by detection of the first arriving PA pulses at each detector location. Prior to the backprojection, the data is ramp filtered, low-pass filtered, and subsequently Hilbert transformed. Next the measured intensities of the US detector are backprojected to discrete points in the PAUS map and back to the epidermis locations according to two different algorithms.. The US waves are considered in the first approximation, to propagate in straight lines and we disregard ray-bending, multiple scattering, aperture weighting and US attenuation. Two variations of the PAUS reconstruction algorithm were tested. The basic PAUS Algorithm (PAUS-1) is based on the synthetic aperture method. Off-line each virtual pulser is considered active at a time; the scattered US from reflection at the bone surface is measured by 32 elements on either side of the acoustic axis of the pulser position. The PAUS images are constructed by the backprojection of the time traces from the 64 elements back to the chosen virtual pulser (Figure 2(c)).

This process is repeated in turn for all the pulsers along the epidermis. In modern probes with increasing array sizes and/or in CT imaging with many projections this can take a considerable amount of system resources. Therefore we also test Algorithm 2 (PAUS-2) in which we consider one pulser/receiver pair at a time, this pair is situated radially from the tomograph center. In other words the pair lie on the line towards the center of rotation. (Figure 2(d)). By considering one pair at a time one-by-one around the object, the aperture is synthesized. This is different from PAUS-1 where the entire receiver array is used per pulser, for all pulsers. This pulser-receiver line in PAUS-2 is approximately perpendicular to the epidermis in most cases due to the circular symmetry of both the probe and finger. This method is highly selective for surfaces of acoustic reflectors that run parallel to the skin surface, which is the case for bone surfaces running close beneath the dorsal side of the finger. This is in contrast to the palmar side where, along most of the finger length, thick tendons are present and the bone surface is relatively flat.

All PA images are plotted in a linear grayscale. The PAUS images are plotted using a logarithmic scale for which a suitable threshold level is manually chosen to visualize the reflectors.

## 3   Results

**Phantom experiments**

*Bone delineation*

Figure 3(a-c) show the PA images of the phantoms P1, P2 and P3 carrying B1, B2 and B3 respectively. The circular shape of the absorbing ink layer at the surface of all agar gel phantoms is clearly visible. In unknown cases, one would mistakenly identify the intensities inside the phantoms to be optical absorbers. However, it is known that these are artifacts caused by acoustic reflections of the surface signals or PAUS signals on the bone-simulating Delrin® samples. These artifacts are visible as a circle in phantom P1, as two half-circles in P2 (Figure 3(b)), or as a three half-circles in phantom P3 (Figure 3(c)). Each flat or round side in a phantom results in a circular artifact with a different radius of curvature. From the PA images of each phantom the location of the ink layer was detected and defined as a collection of discrete PAUS sources; the images were reconstructed using the two proposed algorithms.

*Algorithm 1*

Figure 3(d-f) show the PAUS images constructed by Algorithm 1, and labeled PAUS-1 for each case. The dark regions that are delineated with white lines/areas are the acoustic reflectors. The region outside the reflectors is brighter than the interior of the rods themselves due to backprojection artifacts. It is clear that in the images the shapes of the acoustic reflectors can be identified in comparison with Figures 3(j-l) showing the actual shapes embedded. Especially with curved surfaces, which run roughly concentric to the finger phantom surface, very clear delineation of the reflector surface is obtained. In the case of a flat surface, a striking high-intensity region appears concentrated at the central part of this surface (compare Figure 3e with k for P2, and Figure 3f with l for P3). Furthermore, the contrast seems lower in the corners of the flat bone surfaces than the middle parts in phantom P2 and P3.

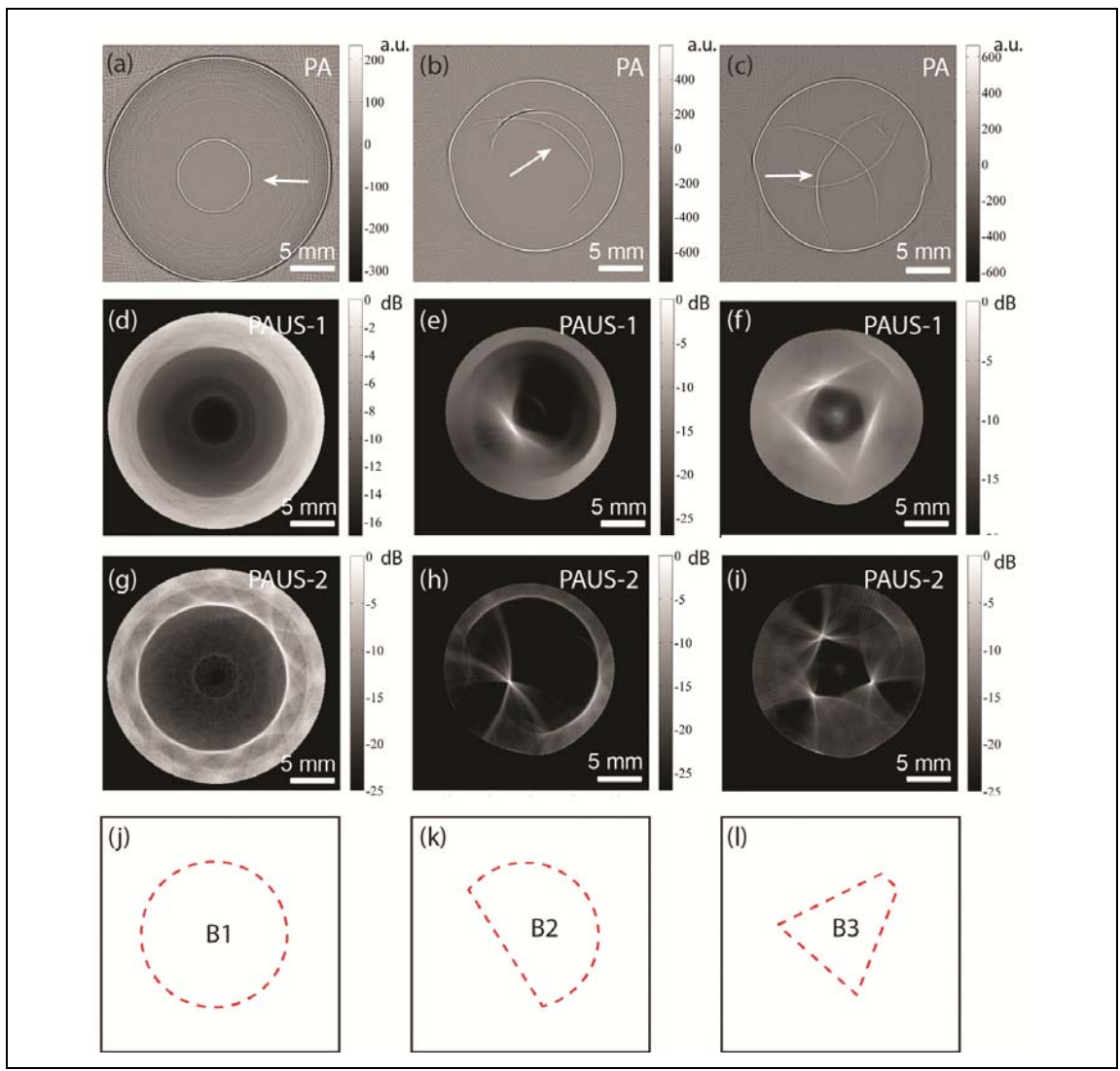

**Figure 3.** (a-c) PA images, (d-f) PAUS images reconstructed with Algorithm 1, and (g-i) PAUS images reconstructed with Algorithm 2 of the lathe turned circular bone-simulating rod (B1), partly circular rod with one flat side, and trapezium shaped rod, respectively. (a-c) clearly show an artifact (solid arrows) for each of the sides of the hidden acoustic reflectors. (d-f) PAUS images made by Algorithm 1 clearly delineate the surfaces of the acoustic reflectors from the curved surfaces, with partial delineation of the flat surfaces. (g-i) PAUS images reconstructed by Algorithm 2 show an improved contrast for circularly shaped acoustic reflectors that lie parallel or are concentric to the absorbing phantom surfaces. The visualization of the flat surfaces however is poor, with a focusing artifact at the centers of the flat surfaces. PAUS = photoacoustically induced ultrasound; PA = photoacoustic.

*Algorithm 2*

The images in Figure 3(g) to (i) show the PAUS images that were reconstructed by Algorithm 2 (PAUS-2). Once again, the dark regions that are delineated with white lines/areas are the acoustic reflectors. With the use of this algorithm, it is clearly observed that the images show excellent contrast in delineating the curved reflector surface compared with PAUS-1. (Compare the results for P2 and P3.) However, the flat reflector surfaces of P2 and P3 are poorly delineated compared with PAUS-1. The surfaces are only partially visible, with only the center parts of these flat surfaces resolved in the images.

In both algorithms, in the cases of a flat bone surface, high intensities are clearly observed at the center while the corners of the flat sides are not resolved. The reason for this is that the PA signals are generated from curved ink and finger surfaces, which produce a natural geometrical focusing of the ultrasound towards the center of the phantom and onto the middle of the flat surface. It must be admitted though that the occurrence of a perfectly flat surface in *in vivo* situations is not realistic. Bones on the dorsal side of fingers run close to and approximately parallel to the skin surface. At the palmar side of the finger the bone is not parallel to the skin due to the presence of a very thick flexor tendon. However, since a large part of the bone surface is concentric with the skin surface, PAUS-2 is the more appropriate algorithm for bone surface delineation and for that reason also to detect the joint space for *in vivo* applications.

*Joint space detection*

An important goal of the PAUS method is to identify the location of the joint space in the finger from PA data. To test the ability of the method in doing so, phantoms P1 and P4 were developed with joint space mimicking geometries.

P1 uses structure B1 to simulate bone which is circularly symmetric and follows the phantom surface in the imaging plane, but is sharply constricted over a length of 3 mm at a certain position along its length. (See Figure 2(b) left) The PA images in Figure 4(a-b) clearly show artifacts arising from the acoustic reflections of the surface signals at surfaces of B1 at (a) the joint space, and at (b) after the joint space. The PAUS-2 method faithfully reconstructs the shapes and sizes of B1 at the respective locations. A complete set of 60 stacked and interpolated slices is shown in Figure 4 (e-f). The PAUS-2 image clearly depicts the surfaces and it is possible to identify the constriction (J). The image shows the best delineation for bone simulating surfaces parallel to and concentric with the phantom surfaces. The delineation of the bone surface is however not strong at all positions along the cross-sectional image. At positions with slopes along the longitudinal axis, the intensity of the reflection signals decreases with increasing angles (dashed arrows). These weak reflections are caused by non-perpendicular specular reflections directed outside the imaging plane. Due to the weakly reflecting slopes the start and the end of the 3 mm joint space are more difficult to visualize. The reflections at the joint space, between the two slopes, are stronger. The reflecting regions at the start and end of the phantom (double headed arrows) are water filled holes in the rod. These regions inside the scattering rod are still visible because part of the US is transmitted instead of reflected.

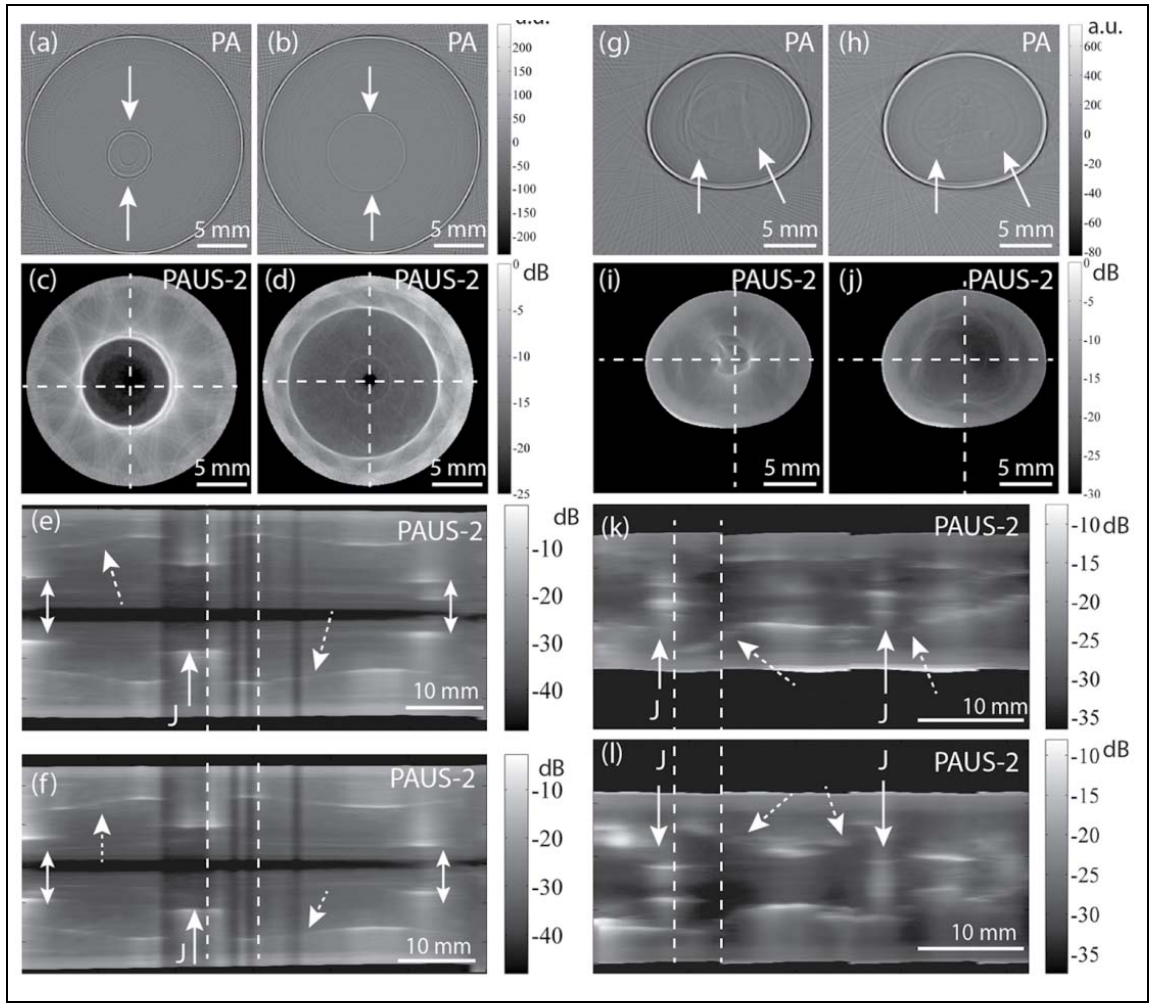

**Figure 4.** Photoacoustic and PAUS-2 images of phantoms P1 and P4. (a-b) Photoacoustic images and (c-d) PAUS-2 images of P1 reconstructed with Algorithm 2. (e-f) Two perpendicular cross-sectional PAUS-2 images over the length of P1 obtained from 60 stacked PAUS-2 slices. (g-h) Photoacoustic images and (i-j) PAUS-2 images of P4 reconstructed with Algorithm 2. (k-l) Sagittal (k) and coronal/dorsal (l) PAUS-2 cross-sectional images of P4 obtained from 60 stacked PAUS slices. Delineation of the surfaces of both P1 and P4 where successful. Small slopes and flat/parallel surfaces are delineated better than surfaces with steep slopes. The intensity is lower at these locations (dashed arrows). Both phantoms show the joint spaces (J). The cylindrical shaped rods between the "bones" of P4 show strong perpendicular reflections (solid arrows in k-l). PAUS = photoacoustically induced ultrasound; PA = photoacoustic.

Figure 4 (g-l) show PA, PAUS-2, and cross-section PAUS-2 images along the length of phantom P4 containing the realistic finger bone model. The stacked images show the contours of the two joint spaces and three finger bones in both the sagittal and coronal cross sections of the skeleton model. The sections of the bone running parallel to the skin are excellently depicted. Similar to phantom P1 the delineation of the bones is weaker at the location next to the two joint spaces due to the steepness of the surface slopes (dashed arrows). The joint space (J) itself was however well visible due to the presence of the small cylindrical shaped rods between the bones which show strong perpendicular specular reflections.

### *In vivo* measurement

The sliced PA, sliced PAUS-2 and stacked PAUS-2 images of a dark skinned human finger are shown in Figure 5. In the PA images in Figure 5(a-b) several details are observed such as the epidermis (circular circumference), some weak blood vessel structures, and more towards the center, artifacts (solid arrows) due to acoustic reflections on the bone. These reflections are primarily present at the dorsal side of the finger because of the higher melanin concentration in the dorsal epidermis and because of the small distance between the epidermis and the bone. The blood vessels are not very well visible because of the high absorption of light by the dark skin at 700 nm.

Figure 5 (c-d) show the transversal PAUS-2 images that correspond to the PA images in the same plane as Figure 5(a-b). These images show bone boundaries that were reconstructed from reflected PA signals generated primarily from the dorsal side of the finger. Delineation of the palmar side of the finger is only partially obtained due to two reasons. Firstly as discussed previously, the method is less successful when the bone surface is relatively flat and not concentric with the skin surface. Secondly, the finger has thick tendon structure at the palmar side which attenuates the signals stronger than at the dorsal side.

The stacked PAUS images of the finger in Figure 5(e-f) shows good delineation of the bone longitudinally along the finger on the dorsal side. Consequently, the contour of the proximal interphalangeal joint space is clearly visualized on the dorsal side, while there is uncertainty in identifying the space on the palmar side.

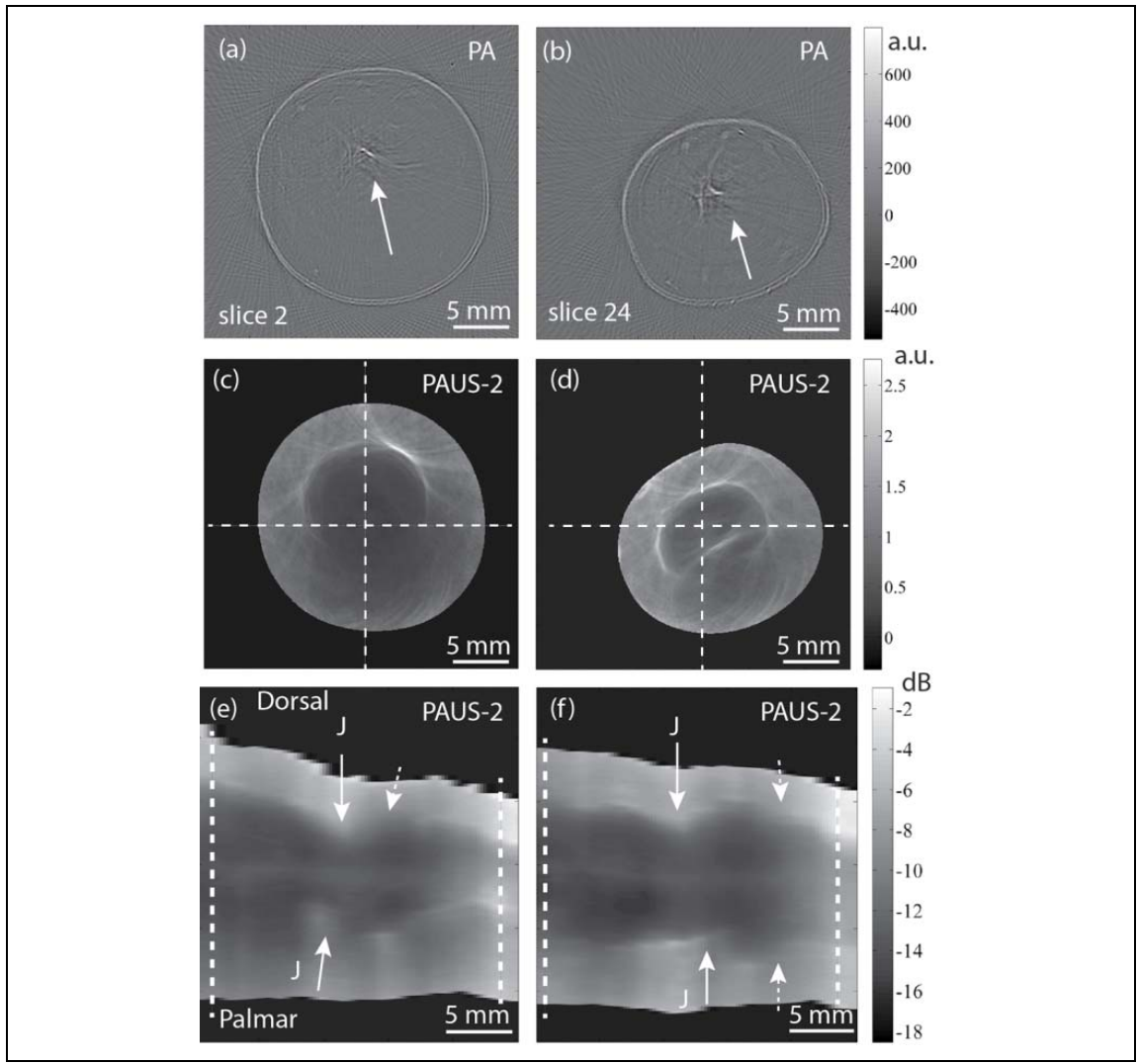

**Figure 5.** In vivo results from a healthy human index finger with a dark skin. (a-b) Transversal OA slices showing epidermis and acoustic reflections (solid arrows) from epidermal signal on the bone surface. (c-d) Transversal PAUS images showing the delineation of bone. (e-f) Sagittal (e) and coronal (f) cross-sections over the length of the proximal interphalangeal joint made by stacking and interpolation of 25 PAUS slices. The solid arrows indicate the location of the joint space based on the change in diameter and loss of US reflections. PAUS = photoacoustically induced ultrasound; US = ultrasound; PA = photoacoustic.

## 4 Discussion

PA imaging visualizes distribution of optical absorption in tissue such as that from hemoglobin and melanin, with high spatial resolutions. The method is very versatile in imaging pathologies associated with vascular abnormalities such as cancer including that of the breast[16-19] and skin[20, 21]. Inflammatory diseases such as rheumatoid arthritis are also associated with vascular malformations such as angiogenesis, and much interest is being shown in the imaging of the phalanges in the region of the joint space to identify the synovium as a surrogate marker of the disease[12, 13, 22].

Especially in imaging the finger with the presence of tendons, cartilage and bone structures it is advantageous to incorporate means to perform simultaneous US imaging. Since the detection hardware of PA imaging typically uses piezoelectric devices, and these can also be used for US transmission, it is logical to develop hybrid imaging using the two modalities. However, US imaging requires high voltage electronics and multiplexers for generation, and the transducers typically have less bandwidth (<60%) than those used in PA imaging. There have been several approaches to use laser-induced US from passive element sources in PA imagers to perform hybrid imaging, where both US transmission and US reflection imaging is performed.[15, 23-29] Hitherto, the passive elements have been strong absorbers introduced into the imaging system exogenously.

In this work, we show that for the specific case of phalanx imaging, we can use strong signals from the skin as endogenous US transmitters to get more information from the bone structures. In this approach PA signals from the skin are considered as a collection of virtual US transmitters, and a form of US reflection imaging is performed that allows the surfaces of the finger's bone to be delineated. This PAUS reflects off bone surfaces and is detected by the US detector.

Two algorithms for PAUS imaging were used, one algorithm called PAUS-1 based on the sector synthetic aperture method, and the second PAUS-2 which only projects PAUS signals back from specular reflecting surfaces due to narrowed receive aperture, arising from the use of a single receiver per pulser. Both algorithms proved successful in delineating the acoustic reflectors. PAUS-1 resolves complex boundaries more accurately, but with lower contrast than PAUS-2 does with bone surfaces concentric with the skin surface. The lower contrast of PAUS-1 can be attributed to the fact that it considers the PAUS sources as point sources while these are highly directional due to focusing of the acoustic waves by the curvature of the finger. Many backprojected lines from side angles contribute to the image while there is less signal present at these angles. PAUS-2 is superior in imaging cylindrically shaped phantoms containing acoustic reflectors parallel to the ink surface, which approximates the finger situation better.

For the phantoms and *in vivo* datasets, we have been successful in delineating the location of the bone surfaces, and consequently the locations of artificial and real joint spaces. The phantom data shows that weak slopes along the length of the bone-simulating structures are visible, but at a much lower intensity, as a result of the non-perpendicular angle of incidence with the surface of the acoustic reflector. Although the accuracy of the delineation is not as rigorously accurate as in conventional ultrasound imaging, it does allow the visualization of the approximate bone surface location. Overlaying the image of the bone surface onto the PA image would therefore provide a more accurate depiction of the situation, and permit better interpretation of the images to distinguish true PA sources from artifacts. Subsequent work will focus on the reduction of reflection artifacts by removal of parts of the PA signal traces that cause reflection artifacts.

Currently, the method is mainly applicable at the dorsal side of the finger. At the palmar side of the finger the absorption of the epidermis is much lower and generates weak PA signals. This, combined with the high attenuation of the ultrasound by the thick flexor tendon between the bone and the skin surface, will result in an overall reduced amount of detected signal. Washable dyes might be used to increase the strength of the PA signal from the skin surface.

Artifacts in the PAUS images caused by blood vessel signals were currently low because of the chosen wavelength and because of the high concentration of melanin in the subject's epidermis. A situation could be thought of, where imaging with at lower wavelengths could be done in addition to standard PA imaging at NIR wavelengths. The use of green light (532 nm) would increase surface absorption and result in stronger PAUS signals for US reflection imaging.

In our case we considered the skin to be relatively smooth in the imaging plane, with wrinkles along the longitudinal axis of the finger. The method is also limited to highly scattering materials such as bone, while tendons and ligaments have not been observed.

The implemented algorithms are basic backprojection algorithms, which are approximate methods. In any case improvements in backprojection algorithms can be made by taking into account the directionality of the focused propagating PAUS waves and including the receive aperture of the ultrasound detectors.[30] This will allow visualization of the flat surfaces with higher contrast.

Due to the large distance between the detector and the skin surface we selected the speed-of-sound (SOS) of water at room temperature (1482 m/s) for all PAUS reconstructions. This is accurate enough for the agar based phantom measurements, but the actual SOS of tissue in the finger varies from 1430 m/s (fat) to 1600 m/s.[31] Implementing a correction for SOS differences will result in more accurately estimated diameters of the bone surfaces for the *in vivo* measurement.

# 5  Conclusion

In photoacoustic computed tomography of the finger, artifacts are produced when strong signals from the epidermis reflect off bone surfaces and are detected. The method which we have presented identifies the bone surfaces and helps to distinguish reflection artifacts from photoacoustic sources. To address this problem, We we treat the epidermis in a the photoacoustic imager as consisting of virtual sources of ultrasound, and treat the detected signals as those in ultrasound reflection imaging. Backprojecting the detected signals to the pulser location allows the delineation of the bone surfaces that caused the reflection. The method which we have presented identifies the bone surfaces and helps to distinguish reflection artifacts from photoacoustic sources and improve interpretation of images.

The method also allows provides co-registered ultrasound images to be developed, and permits the depiction of the joint space without the need for additional hardware or measurements. The joint space can help to locate the expected position of the synovial membrane which is important in the photoacoustic imaging of the joints in the context of inflammatory arthritis.

## ACKNOWLEDGEMENTS

We thank Dr. H. J. Bernelot Moens of the Ziekenhuisgroep Twente (ZGT), Hengelo and Almelo, for the hand bone model B4 used in phantom P4. This research is financially supported by the Netherlands Organization for Health Research and Development (ZonMw) under the program New Medical Devices for Affordable Health; and the High-Tech Health Farm Initiative of the Overijssel Center for Research and Innovation (OCRI). WS and SM are minority shareholders in PA Imaging BV, however the company did not financially support this research